\documentclass[twocolumn,superscriptaddress,prc,showpacs,amsmath,amssymb,letterpaper]{revtex4-1}
\usepackage{graphicx,color}
\usepackage{amssymb}
\usepackage{amsmath}
\usepackage{latexsym}
\usepackage{amsxtra}
\usepackage{amscd}
\usepackage{dcolumn}
\usepackage{bm}
\usepackage[breaklinks]{hyperref}
\usepackage{url}
\usepackage{breakurl}

\graphicspath{{plots/}}
\let\oldsqrt\sqrt
\def\sqrt{\mathpalette\DHLhksqrt}
\def\DHLhksqrt#1#2{\setbox0=\hbox{$#1\oldsqrt{#2\,}$}\dimen0=\ht0
\advance\dimen0-0.2\ht0
\setbox2=\hbox{\vrule height\ht0 depth -\dimen0}%
{\box0\lower0.4pt\box2}}

\newcommand{\gray}{$\gamma$ ray}
\newcommand{\ghray}{$\gamma$-ray}
\newcommand{\grays}{$\gamma$ rays}
\newcommand{\nuc}[2]{$^{#1}$#2}
\newcommand{\cs}{cross section}
\newcommand{\md}{momentum distribution}
\renewcommand{\sf}{spectroscopic factor}
\newcommand{\sfs}{spectroscopic factors}
\begin{document}

\title{Shell structure of \boldmath{$^{43}$}S and collapse of the \boldmath{$N=28$} shell closure}

\author{S.~Momiyama} 
\affiliation{Department of Physics, The University of Tokyo, Hongo, Bunkyo-ku, Tokyo 113-0033, Japan}
\author{K.~Wimmer} 
\affiliation{Department of Physics, The University of Tokyo, Hongo, Bunkyo-ku, Tokyo 113-0033, Japan}
\affiliation{Instituto de Estructura de la Materia, CSIC, E-28006 Madrid, Spain}
\author{D.~Bazin} 
\author{J.~Belarge}
\author{P.~Bender}
\affiliation{National Superconducting Cyclotron Laboratory, Michigan State University, East Lansing, Michigan 48824, USA}
\author{B.~Elman} 
\author{A.~Gade}
\affiliation{National Superconducting Cyclotron Laboratory, Michigan State University, East Lansing, Michigan 48824, USA}
\affiliation{Department of Physics and Astronomy, Michigan State University, East Lansing, Michigan 48824, USA}
\author{K.~W.~Kemper}
\affiliation{Department of Physics, Florida State University, Tallahassee, Florida 32306, USA}
\author{N.~Kitamura}
\affiliation{Center for Nuclear Study, University of Tokyo, Wako, Saitama 351-0198, Japan}
\author{B.~Longfellow} 
\author{E.~Lunderberg}
\affiliation{National Superconducting Cyclotron Laboratory, Michigan State University, East Lansing, Michigan 48824, USA}
\affiliation{Department of Physics and Astronomy, Michigan State University, East Lansing, Michigan 48824, USA}
\author{M.~Niikura}
\affiliation{Department of Physics, The University of Tokyo, Hongo, Bunkyo-ku, Tokyo 113-0033, Japan}
\author{S.~Ota}
\author{P.~Schrock}
\affiliation{Center for Nuclear Study, University of Tokyo, Wako, Saitama 351-0198, Japan}
\author{J.~A.~Tostevin}
\affiliation{Department of Physics, University of Surrey, Guildford, Surrey GU2 7XH, United Kingdom}
\author{D.~Weisshaar}
\affiliation{National Superconducting Cyclotron Laboratory, Michigan State University, East Lansing, Michigan 48824, USA}

\begin{abstract}
  The single-particle structure of the $N=27$ isotones provides insights into the shell evolution of neutron-rich nuclei from the doubly-magic \nuc{48}{Ca} toward the drip line. \nuc{43}{S} was studied employing the one-neutron knockout reaction from a radioactive \nuc{44}{S} beam. Using a combination of prompt and delayed \ghray\ spectroscopy the level structure of \nuc{43}{S} was clarified. Momentum distributions were analyzed and allowed for spin and parity assignments. The deduced spectroscopic factors show that the \nuc{44}{S} ground-state configuration has a strong intruder component. The results were confronted with shell model calculations using two effective interactions. General agreement was found between the calculations, but strong population of states originating from the removal of neutrons from the $2p_{3/2}$ orbital in the experiment indicates that the breakdown of the $N=28$ magic number is more rapid than the theoretical calculations suggest.
\end{abstract}

\date{\today}
\pacs{
}
\maketitle
\section{Introduction}\label{sec:intro}

The emergence of shell closures or their disappearance in exotic nuclei has been one of the main interests of the nuclear structure community since the advent of radioactive beam facilities. Islands of inversion and shape coexistence have been associated with the disappearance of the classical shell closures on the neutron-rich side of the valley of stability~\cite{gade16b}. In particular, the $N=28$ shell closure, arising in a harmonic oscillator plus spin-orbit mean field, has recently attracted much interest~\cite{sorlin13}. Below the doubly magic nucleus \nuc{48}{Ca} with 20 protons and 28 neutrons, the $N=28$ nuclei show a variety of interesting features. Mass measurements~\cite{meisel15}, transfer~\cite{gaudefroy08}, and nucleon knockout reactions~\cite{gade05} support a strong $N=28$ shell closure in \nuc{46}{Ar}. Measurements of the reduced transition probability, $B(E2)$, find a low degree of collectivity~\cite{scheit96,gade03}, while shell model calculations show enhanced collectivity at the shell closure. This discrepancy is yet to be resolved. In \nuc{44}{S}, the measurement of a large $B(E2)$ value and its comparison to theoretical calculations suggested a vibrational character of this nucleus~\cite{glasmacher97}. The lowering of the excited $0^+_2$ state from 3695~keV in \nuc{46}{Ar}~\cite{nowak16} to 1365~keV~\cite{grevy05} in \nuc{44}{S} indicates the onset of shape coexistence and a rapid weakening of the $N=28$ shell closure. The measured $E0$ strength between the \nuc{44}{S} $0^+$ states was interpreted as arising from the substantial mixing of spherical and prolate configurations~\cite{force10}.
Theoretical calculations of the potential energy surface using the symmetry-conserving configuration mixing method and the Gogny D1S interaction do not show distinct minima characteristic of shape coexistence and rather suggest configuration mixing~\cite{rodriguez11}. Later refinements of the theory and extended calculations find that the ground state of \nuc{44}{S} has a collective wave function which is extended in the ($\beta,\gamma$) plane while the excited $0^+_2$ is prolate, yet $\gamma$-soft~\cite{egido16}.
Shell model calculations using a newly derived SDPF-MU interaction~\cite{utsuno12} suggest that the evolution of collectivity along $N=28$ is governed by the proton-neutron tensor force~\cite{utsuno15}. Here, the potential energy surface exhibits a minimum on the prolate side.

The \nuc{42}{Si} nucleus is well deformed, it exhibits a low excitation energy for the first $2^+$ state~\cite{bastin07} and a large $R_{4/2}$ ratio~\cite{takeuchi12}. Calculations with the SDPF-MU interaction predict the ground state of \nuc{42}{Si} to be strongly oblate deformed~\cite{utsuno15}. Detailed spectroscopy of \nuc{42}{Si}, however, questioned the $4^+$ assignment of Ref.~\cite{takeuchi12} and proposed an excited $0^+_2$ state based on the observed population cross section~\cite{gade19}.
Approaching the drip-line~\cite{ahn19}, the last $N=28$ nucleus with excited states known is \nuc{40}{Mg}~\cite{crawford19}. The measured two-proton removal cross sections along the $N=28$ isotones~\cite{takeuchi12,crawford14} were interpreted as showing a change of the ground state deformation from prolate in \nuc{44}{S} to oblate for \nuc{42}{Si}, and back to prolate at \nuc{40}{Mg}.

Turning to the even-odd $N=27$ nuclei, \nuc{43}{S} has attracted special attention, both from the theoretical and experimental side. In \nuc{45}{Ar} the ground state is $7/2^-$, as expected from the normal orbital filling. A low-lying $J^\pi = 3/2^-$ state with a rather long lifetime~\cite{dombradi03} is strongly populated in the $(d,p)$ reaction adding a neutron to \nuc{44}{Ar}~\cite{gaudefroy08} and very weakly in the neutron removal reaction~\cite{gade05}. This confirms the vacancy of the $2p_{3/2}$ orbital in both \nuc{44,46}Ar and the existence of a shell closure at $N=28$. In \nuc{43}{S}, an isomeric state with a lifetime of 478(48)~ns was found at 319~keV~\cite{sarazin00}. Based on the comparison with shell model calculations the isomeric state was assigned spin and parity $7/2^-$ and a level inversion compared to \nuc{45}{Ar} was proposed. A measurement of the magnetic moment firmly assigned $J^\pi = 7/2^-$ to the isomeric state and, because its lifetime is only compatible with an $E2$ transition, the ground state was inferred as $J^\pi = 3/2^-$~\cite{gaudefroy09}. The spherical nature of the $7/2^-$ isomeric state was questioned and the spectroscopic quadrupole moment, determined to be $|Q_\text{s}|=23(3)$~efm$^2$, was significantly larger than the expectation for a single hole in the $1f_{7/2}$ orbital. While the state cannot be considered spherical, shell model calculations do not predict a band structure built upon the isomeric state~\cite{chevrier12}. These results triggered various theoretical discussions. Antisymmetrized molecular dynamics (AMD) calculations indicate that the $7/2^-$ isomer might be triaxial, and that bands of prolate, oblate, and triaxial nature coexist at low excitation energy~\cite{kimura13}. The gap between neutron single-particle levels originating from the spherical $1f_{7/2}$ and $2p_{3/2}$ orbitals reduces as a function of the deformation parameter $\beta_2$; the two orbitals cross around a prolate deformation with $\beta_2 \approx 0.2$ and the $N=28$ shell gap disappears. A state at around 940~keV observed in a Coulomb excitation measurement~\cite{ibbotson99} is suggested as the $7/2^-$ member of the prolate $K^\pi=1/2^-$ ground state band with a negative decoupling parameter. An oblate band built on the $3/2^-_2$ state is also predicted. A shell model study exploiting quadrupole rotational invariants came to similar conclusions~\cite{chevrier14}. The calculations based on the SDPF-U effective interaction~\cite{nowacki09} predict a third, prolate band with a dominant 2p-2h $(1f_{7/2})^{-3}(2p_{3/2})^2$ configuration. Calculations using the SDPF-MU effective interaction~\cite{utsuno12} and the variation after angular-momentum projection method show that the ground state and the isomeric state are dominated by $K=1/2$ and $7/2$ and the isomeric nature is explained by the $K$ forbiddeness of the decay~\cite{utsuno15}. This interpretation also explains the occurrence of the long-lived $0^+_2$ and $4^+_1$ states in \nuc{44}{S}~\cite{grevy05,force10,santiago11,parker17}. 

Spectroscopic information on states in \nuc{43}{S} beyond the ground and isomeric state was obtained from nucleon removal reactions, however, placement in the level scheme proved difficult because of the presence of the isomeric state~\cite{riley09}. Most recently, excited state lifetimes in \nuc{43}{S} were measured. Using the proton knockout reaction from \nuc{44}{Cl} several states were populated~\cite{mijatovic18}. The level ordering was reversed compared to the earlier study~\cite{riley09}. It should be noted that the level scheme and the interpretation of Ref.~\cite{mijatovic18} are at variance with the results presented here. In the present work, the neutron knockout reaction is measured with the additional capability to distinguish between decays to the isomer and to the ground state. 

In the present paper, we report on the measurement of the single-particle structure of \nuc{43}{S} using the one-neutron knockout reaction from a fast radioactive \nuc{44}{S} beam. The combination of prompt and delayed spectroscopy allowed for an unambiguous construction of the level scheme and the extraction of spectroscopic factors using reaction model calculations. The results suggest an intruder-dominated configuration in the ground state of \nuc{44}{S}.

\section{Experiment}\label{sec:exp}
The experiment was performed at the Coupled Cyclotron Facility of the National Superconducting Cyclotron Laboratory at Michigan State University~\cite{gade16}. The secondary \nuc{44}{S} beam was produced by projectile fragmentation of a 140~$A$MeV \nuc{48}{Ca} primary beam on a 705~mg/cm$^2$ \nuc{9}{Be} production target located at the entrance of the A1900 separator~\cite{morrissey03}. The beam particles were identified by their time-of-flight on an event-by-event basis. The secondary beam was separated and transported to a 376(4)~mg/cm$^2$ \nuc{9}{Be} secondary target located at the pivot point of the S800 spectrograph~\cite{bazin03}. The momentum acceptance of the A1900 separator was set to 1\%, resulting in a mid-target energy of 93.7~$A$MeV and an average \nuc{44}{S} intensity and purity of about 1900~pps and 98(1)\%, respectively. 

\begin{figure}[h]
\centering
\includegraphics[width=\columnwidth]{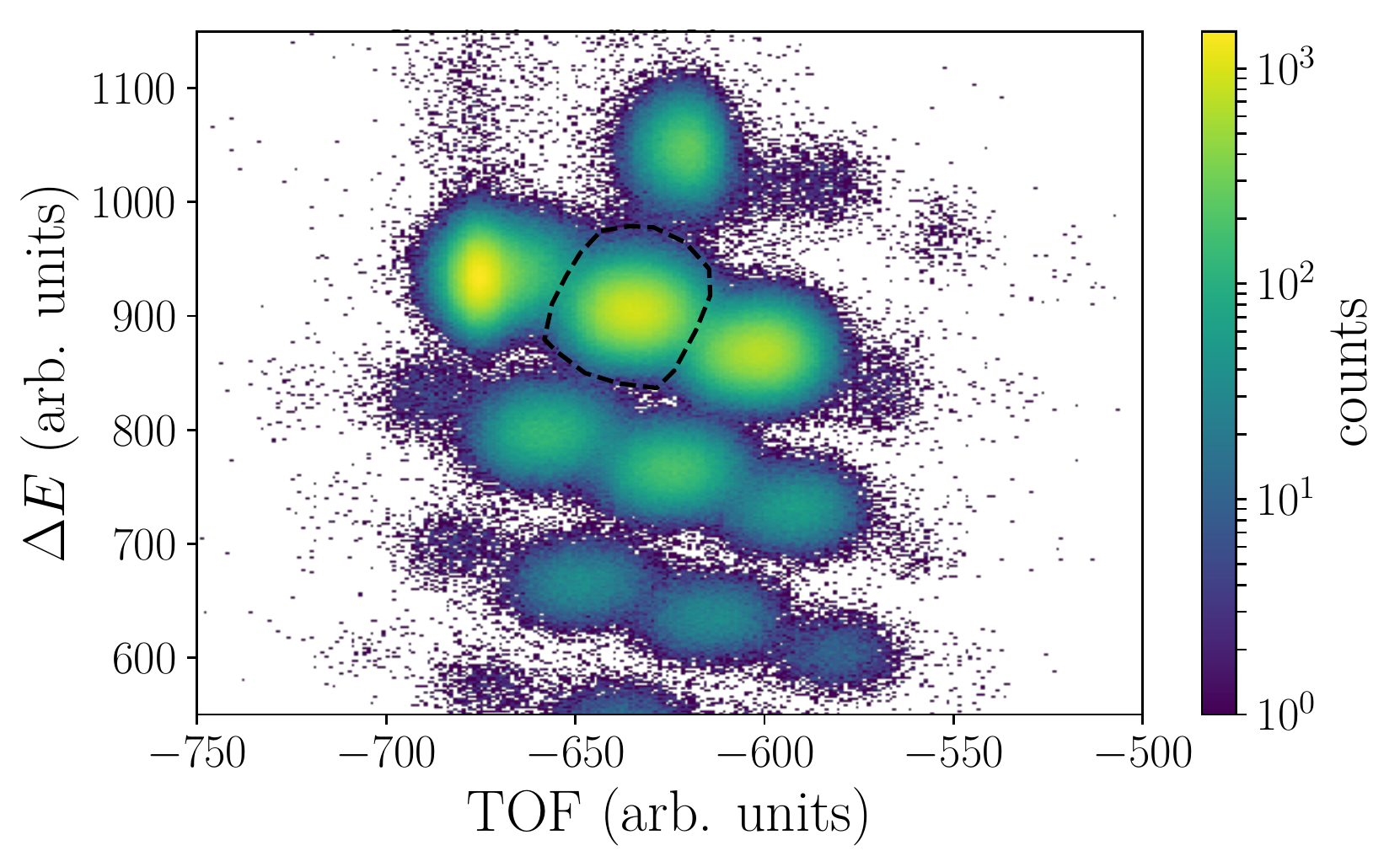}
\caption{Particle-identification plot of reaction residues detected in the S800 spectrograph. A gate on incoming \nuc{44}{S} ions is applied. The dashed line is the outgoing \nuc{43}{S} gate for the further analysis.}
\label{fig:pid}
\end{figure}

The reaction residues were analyzed and identified in the S800 spectrograph~\cite{bazin03} as shown in Fig.~\ref{fig:pid}.
Particle identification was achieved by measuring the energy loss in an ionization chamber ($\Delta E$) in the focal plane of the S800 spectrograph and the time-of-flight (TOF) between two plastic scintillators located before the target and in the focal plane, respectively. Positions and angles of reaction residues at the end of the S800 spectrograph were measured by two cathode-readout drift chambers (CRDC) and traced back to the secondary target by using the ion optics code COSY Infinity~\cite{berz93}. This allowed the determination of the non-dispersive position and the momentum vector at the secondary target. In order to improve the resolution for the momentum transfer, a parallel plate avalanche counter (PPAC) was placed at the intermediate image plane upstream of the target. Here, the dispersive position is correlated with the momentum of the projectile, and the momentum of the incoming projectile can thus be obtained. The momentum resolution for the incoming beam with the PPAC position correction was deduced as 0.052~GeV/c.

The secondary target was surrounded by the Gamma Ray Energy Tracking In-beam Nuclear Array (GRETINA)~\cite{paschalis13,weisshaar17}. A GRETINA module consists of four high-purity germanium crystals, each 36-fold segmented.
In the present experiment, four detector modules were placed at $58^\circ$ with respect to the beam axis and four were placed at 90$^\circ$. The signals were digitized and an online pulse-shape analysis algorithm allowed for the determination of \ghray\ interaction points with energy and position information. It was assumed that the hit with the largest energy deposition was the first interaction, and its position was used for the Doppler correction. The \ghray\ position information was also used in the tracking analysis, where \ghray\ interactions were added together when the difference between their emission angle with respect to the target position was less than $25^\circ$. This add-back analysis was adopted for the $\gamma$-$\gamma$ coincidence analysis and for extracting the exclusive parallel momentum distributions. The energy and efficiency calibration of GRETINA was done with standard radiation sources and the deviation from literature values were deduced to be less than 1~keV. The efficiency of the whole array was measured to be 5.9\% at 1~MeV. 
The \ghray\ yields were determined from a fit of simulated response functions to the \ghray\ energy spectrum. The experimental setup was implemented in a GEANT4 simulation~\cite{agostinelli03} including the experimentally determined thresholds and resolutions of each individual Ge crystal. In the $\chi^2$ fit, the \ghray\ energies and intensities were individually varied to reproduce the measured spectrum.

Finally, the reaction residues were implanted into a 6.35~mm thick Al plate at the back of the focal plane of the S800 spectrograph. Delayed \grays\ emitted from the decay of isomeric states were detected in IsoTagger~\cite{wimmer15} consisting of 32 CsI(Na) detectors. This allowed construction of the level scheme above the 320~keV isomeric state in \nuc{43}{S} for the first time and deduction of the population cross sections for all final states. The energy and efficiency calibration of IsoTagger was performed with a standard \nuc{88}{Y} source. The efficiency at 898~keV was measured to be 8.3\%.

The \nuc{44}{S} nucleus has an isomeric $0^+_2$ state at 1365~keV with a 2.619(26)~$\mu{\rm s}$ half-life~\cite{grevy05,force10}. The beam can thus reach the secondary target in an excited state. In addition to the direct $E0$ transition to the ground state this isomeric state also decays to the $2^+_1$ state with a branching ratio of  16.3(13)\%~\cite{force10}. The \ghray\ transition from the $2^+_1$ state to the ground state could have been observed in the IsoTagger, however, no transition at this energy was observed. The isomeric ratio of the $0^+_2$ state in \nuc{44}{S} is thus assumed to be negligible for the extraction of cross sections. 

\section{Results}\label{sec:results}
\begin{figure}[h]
\centering
\includegraphics[width=\columnwidth]{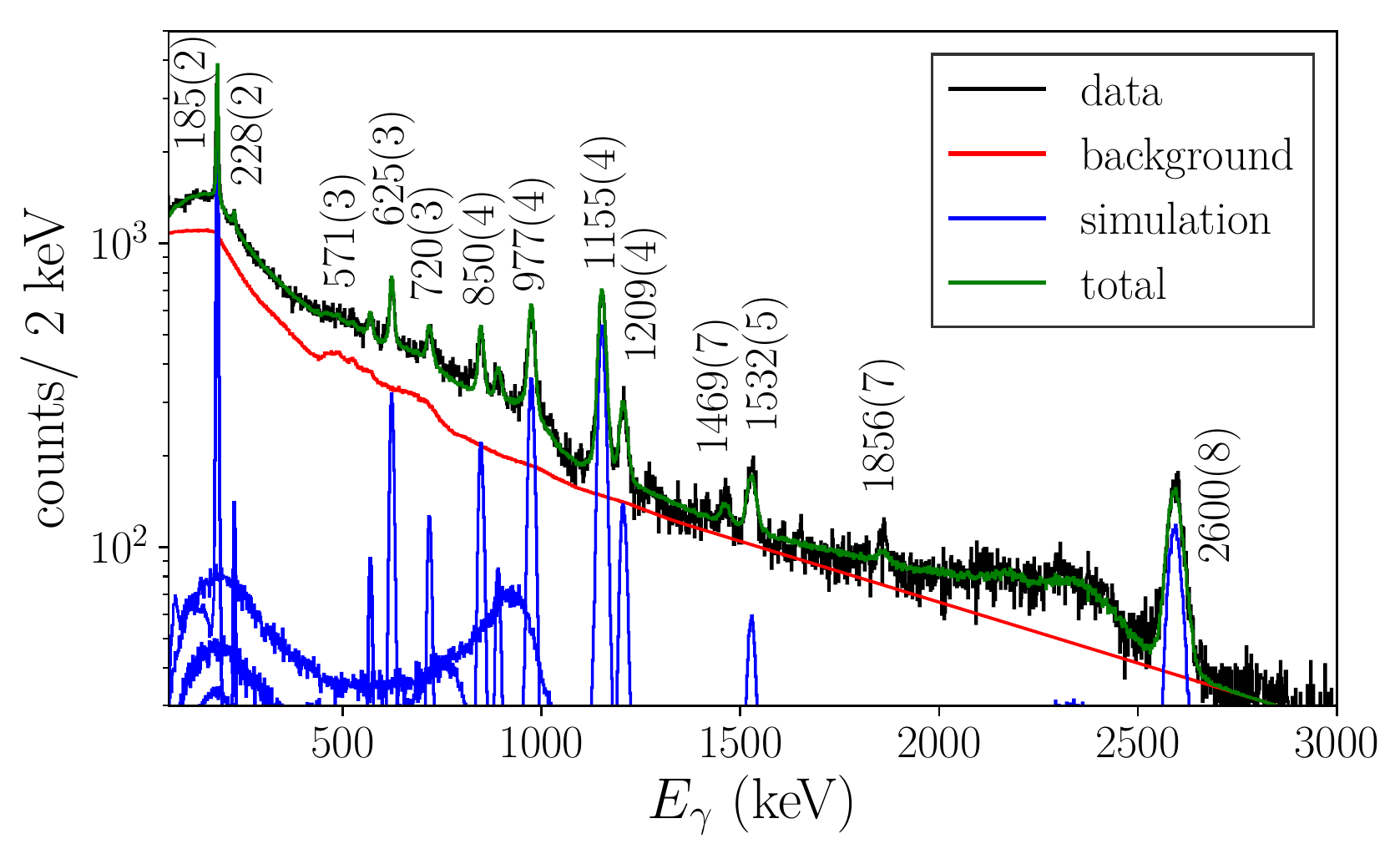}
\caption{Prompt, Doppler-corrected \ghray\ energy spectrum for the one-neutron knockout reaction from \nuc{44}{S} to \nuc{43}{S}. The peaks are labeled with the transition energy and uncertainty in keV. The background around 500~keV includes transitions from neutron-induced reactions on Ge and Al.}
\label{fig:egamdc}
\end{figure}

\begin{figure}[h]
  \centering
\includegraphics[width=\columnwidth]{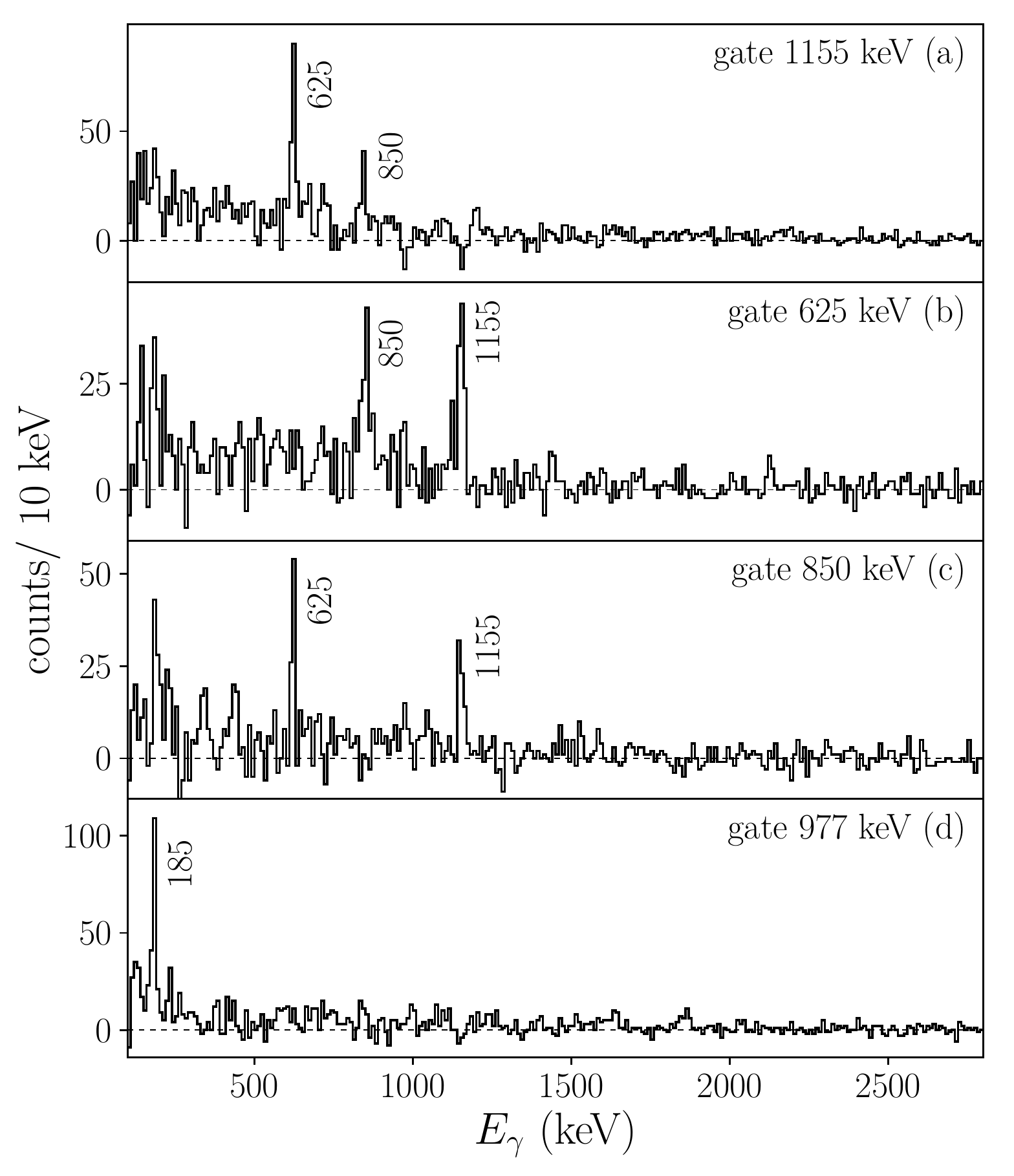}
\caption{Background subtracted $\gamma$-$\gamma$ coincidence spectra measured in GRETINA. Panels (a), (b), (c), and (d) show the spectra gated on the prompt 1155, 625, 850, and 977~keV transitions.}
\label{fig:gamgam}
\end{figure}

Fig.~\ref{fig:egamdc} shows the prompt, Doppler-corrected \ghray\ energy spectrum measured with GRETINA gated on the one-neutron knockout reaction from \nuc{44}{S} to \nuc{43}{S}. 
Most of the previously observed \grays~\cite{riley09,mijatovic18} were confirmed and their energies are shown in Fig.~\ref{fig:egamdc} together with their uncertainty. The transition at 571(3)~keV is newly observed in this work. 
For the error estimation of the \ghray\ energy, the uncertainties of the energy calibration of GRETINA,  the velocity of \nuc{43}{S} for the Doppler correction, and a potential offset of the reaction target location along the beam axis were considered. The individual contributions were, for example for the 2600~keV transition, less than 0.5, 3, and 6~keV.
To deduce the yield of each prompt \gray, a $\chi^2$ fit of the simulated response functions to the experimental spectrum was performed. In this fitting procedure, background \grays\ of neutron-induced reactions with the Ge detectors and surrounding materials were also considered. In the laboratory system clear peaks around 600~keV are observed. The remaining continuous background was modeled as the sum of two exponential functions connected to a linear function in the lower energy region. 
The uncertainties for the \ghray\ yields include, besides the statistical uncertainty, consideration of the deviation of the simulated efficiency from the measured one. This contribution was smaller than 4\% over the whole energy range and thus smaller than the statistical uncertainties. The prompt \ghray\ energies and intensities are compiled in Table~\ref{tab:gamma}. 
\begin{table}
  \caption{\label{tab:gamma} 
    Observed \gray\ energies, efficiency-corrected intensities, and coincidence information for \nuc{43}{S}. The uncertainties of the \gray\ energies include all systematic uncertainties while yields include only the statistical errors. }
  \begin{ruledtabular}
    \begin{tabular}{rrrr}
      energy (keV) & yield/ion (\%) & coincident \grays & level (keV) \\
      \hline
      185(2)& 5.8(3)    &       977 & 1162(4)\\
      228(2)& 0.44(7)   &           &  228(2)\\
      320(2)& 49(3)     &      1532 &  320\\
      571(3)& 0.93(11)  &           & \\
      625(3)& 3.6(2)    & 850, 1155 & 1780(5)\\
      720(3)& 1.8(2)    &           & \\
      850(4)& 3.6(2)    & 625, 1155 & 2628(6)\\
      977(4)& 7.1(4)    &       185 &  977(4)\\
      1155(4)& 13.2(6)  & 625,  850 & 1155(4)\\
      1209(4)& 3.6(2)   &           & 1209(4)\\
      1469(7)& 0.67(13) &           & 2628(6)\\
      1532(5)& 2.2(2)   &       320 & 1854(4)\\
      1856(7)& 0.37(13) &           & 1854(4)\\
      2600(8)& 9.7(5)   &           & 2600(8)\\
    \end{tabular}
  \end{ruledtabular}
\end{table}
Fig.~\ref{fig:gamgam} shows the background subtracted $\gamma$-$\gamma$ coincidence spectra gated on the 1155, 625, 850 and 977~keV transitions. 
The three transitions at 1155, 625, and 850~keV are emitted in cascade and the 977 and 185~keV transitions are in mutual coincidence, but not with any of the other transitions. 
This is in agreement with the level scheme proposed in Ref.~\cite{mijatovic18} with a doublet of states at 1155 and 1162 keV.
The high statistics obtained in the present work makes it possible to determine the order of the \ghray\ transitions in the cascades by the comparison of the measured \ghray\ intensities in Table~\ref{tab:gamma}. These intensities confirm the order of the $850\rightarrow625\rightarrow1155$~keV and $185\rightarrow977$~keV cascades. The latter is opposite to the suggestion of Ref.~\cite{mijatovic18} and thus challenges the result of the very similar lifetimes of the states at 185 and 1162~keV proposed in that work. The present ordering of the cascade is also consistent with earlier measurements of Coulomb excitation~\cite{ibbotson99} assuming that the transition observed around 940~keV corresponds to the 977 peak observed in the present work. In fact, the transition energy is not determined accurately in Ref.~\cite{ibbotson99}, and the observed line could be composed of several transitions within the limited energy resolution. The isobar \nuc{43}{Cl} and the isotone \nuc{45}{Cl} have transitions at 943 and 928~keV which could have contaminated the spectrum. A recent Coulomb excitation experiment confirmed the 977~keV state that is directly excited from the ground state~\cite{longfellow20}. 
The 850~keV transition is placed on top of the 625~keV one, since the former was not observed in the proton removal reaction~\cite{mijatovic18}.
The transition at 1469~keV was placed to feed either the 1155~keV or the 1162~keV state from the 2628~keV state based on the matching energy sum. No coincidences were found for the 1209 and 2600~keV transitions. Based on their intensities, coincidences should have been observed and these transitions are therefore placed as direct ground state decays. The transitions at 228, 571, and 720~keV could not conclusively be placed in the level scheme due to limited statistics. The 228~keV transition is placed as a direct ground state decay from the first excited state at 228~keV, based on the comparison with theoretical calculations (see Section~\ref{sec:discussion}). 

\begin{figure}[h]
\centering
\includegraphics[width=\columnwidth]{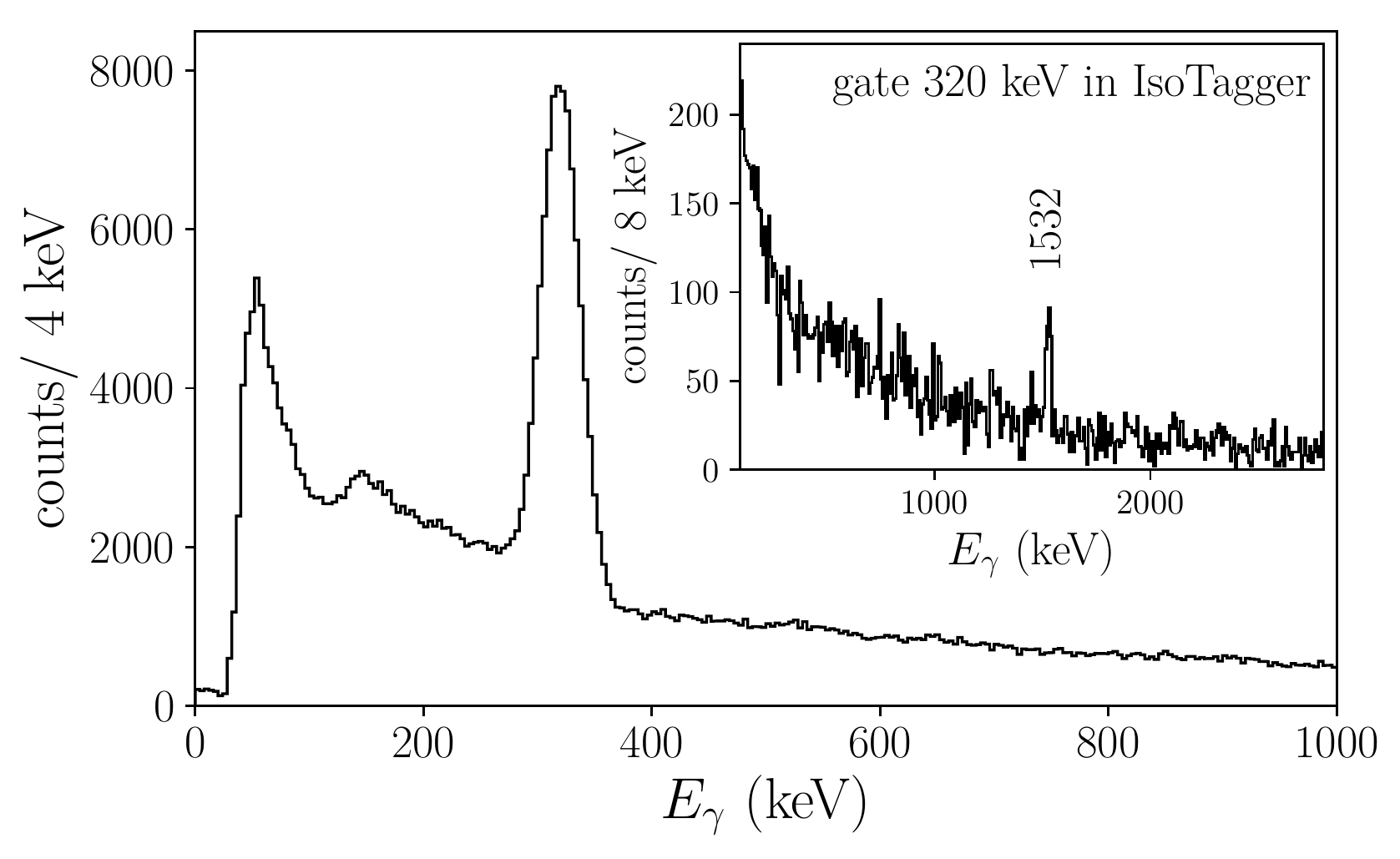}
\caption{Gamma-ray energy spectrum measured by IsoTagger. A gate on \nuc{43}{S} has been applied. The isomeric decay of the 320~keV state is observed. The inset shows the prompt, Doppler-corrected \ghray\ energy spectrum measured with GRETINA gated on the delayed 320~keV transition.}
\label{fig:hodo}
\end{figure}

Fig.~\ref{fig:hodo} shows the \ghray\ energy spectrum measured by the IsoTagger in delayed coincidence with identified \nuc{43}{S} reaction residues.
The decay of the known 320~keV isomeric state~\cite{sarazin00} is observed. The intensity of the 320~keV transition was determined from a $\chi^2$ fit of a simulated response function~\cite{wimmer15} to the spectrum in a similar manner as for the prompt spectrum. The background was modeled as the sum of two exponential functions. The implantation position distribution of the \nuc{43}{S} ions was implemented in the simulation as described in Ref.~\cite{wimmer15}. The position on the stopper plate was taken from the experimental $xy$ distribution measured by the CRDC detectors in the S800 focal plane and extrapolated to the stopper plate. The implantation depth, $z$ coordinate, was estimated by the ATIMA code~\cite{weik} using the experimentally measured energy distribution of \nuc{43}{S} ions. In order to extract the yield, the in-flight decay of the isomer between the secondary reaction target and the stopper plate needed to be taken into account. The half-life of the isomeric state has been previously measured ($T_{1/2} = 478(48)$~ns~\cite{sarazin00}, 415(5)~\cite{gaudefroy09}, and $200^{+140}_{-70}$~ns~\cite{kameda12}). 
In the present experiment, the half-life was determined from the decay curve after implantation. The result of $T_{1/2} = 391(14)$~ns is slightly lower than the most precise value but consistent. Considering the trajectory and the velocity of \nuc{43}{S} behind the secondary target, 79.4(23)\% of the isomeric state initially produced at the target reached the stopper. For the uncertainty estimation on the yield, the deviation of the present half-life from the previous measurement of 415(5)~ns~\cite{gaudefroy09}, the velocity distribution of the \nuc{43}{S} reaction products, and the effect of the uncertainty of the simulated implantation depth on the efficiency of IsoTagger (2\% at 320~keV) were considered.

Fig~\ref{fig:hodo} also shows the prompt \grays\ detected in GRETINA in delayed coincidence with the decay of the 320~keV isomer. The 1532~keV transition is clearly in coincidence with the isomeric transition and the energy sum matches the 1856~keV transition. This establishes a new state at 1854(4)~keV using the weighted average of the energies. Looking for coincidences with the 1532~keV transition in GRETINA does not reveal another \ghray\ transition as a candidate for a transition on top of the isomer.

\begin{figure}[h]
\centering
\includegraphics[width=\columnwidth]{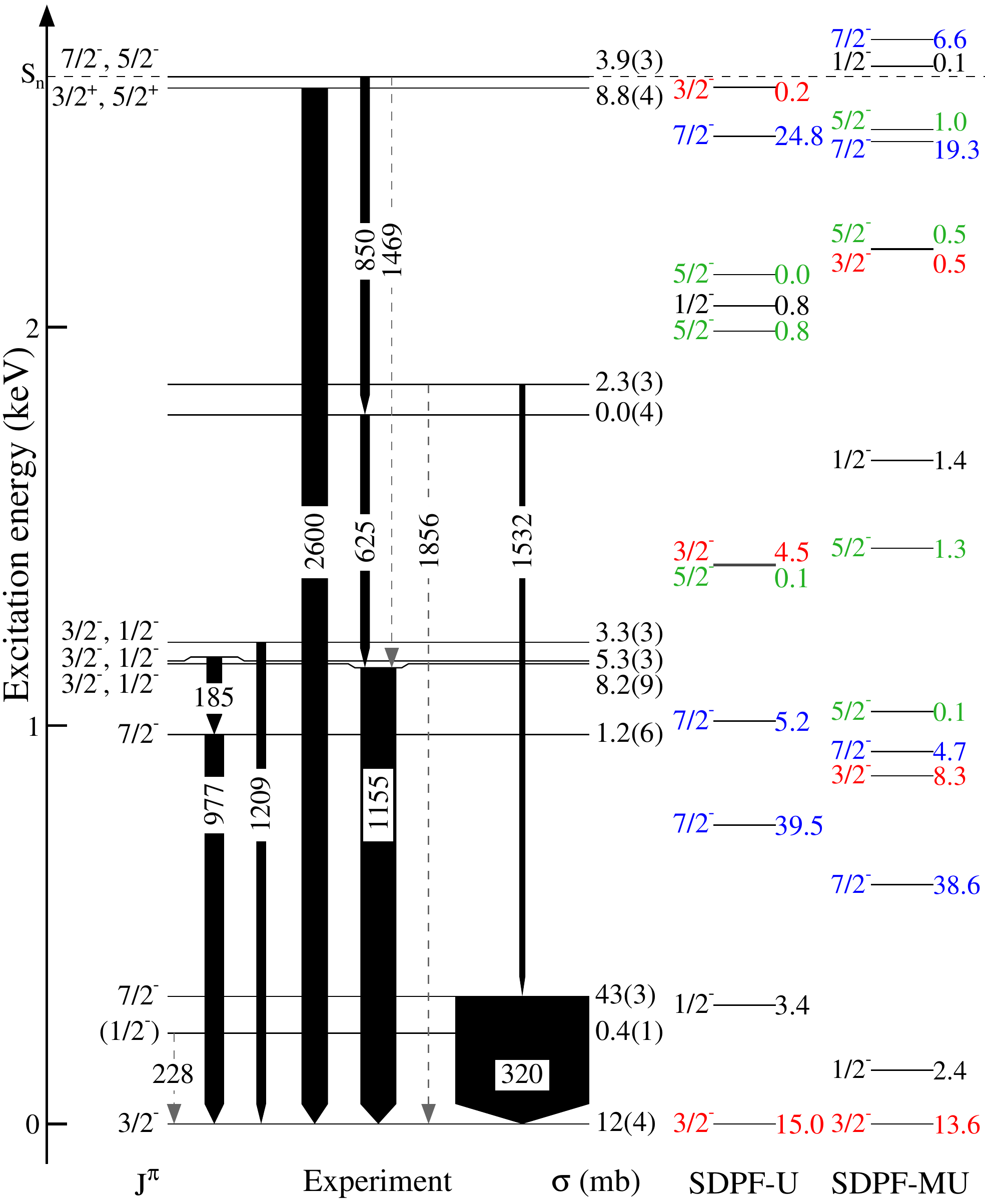}
\caption{Level scheme of \nuc{43}{S} determined from the present experimental results and predicted by shell model calculations. The width of the arrows reflects the measured \ghray\ yields. Gray, dashed transitions are place based on the energy differences of established levels, the 228~keV state is placed based on the comparison to the shell model calculations. The levels are labeled with the spin and parity assignments derived from the measured momentum distributions and the partial cross section for each state (in mb). Spins and parities of predicted $1/2^-$, $3/2^-$, $5/2^-$, and $7/2^-$ states are indicated in black, red, green, and blue, respectively. Theoretical cross sections (in mb) include the calculated spectroscopic factors and the reaction model calculations for the single-particle cross sections (see text for details).}
\label{fig:level}
\end{figure}

The level scheme of \nuc{43}{S}, determined in the present work, is shown in Fig.~\ref{fig:level}.
The order of the transitions of a \ghray\ cascade was determined by comparing the observed yields. The 1469 and the 1856~keV transitions were placed in the level scheme solely based on energy differences. Two states are located close to the neutron separation energy $S_\text{n} = 2629$~keV~\cite{ringle09}. The 2600~keV state decays directly to the ground state, while the 2628~keV state decays via a cascade. The fact that the 2600~keV transition was not observed in the fragmentation reaction of \nuc{45}{Cl}~\cite{riley09} nor in the proton knockout reaction~\cite{mijatovic18} from \nuc{44}{Cl} supports the presence of two different states. The very different momentum distributions (see below) for the 2600 and 2628~keV states further confirm the existence of two close-lying states near the neutron separation energy.

Using the level scheme presented in Fig.~\ref{fig:level} the final-state exclusive cross sections were determined. They are presented in Table~\ref{tab:cs} and Fig.~\ref{fig:level}.
\begin{table}
  \caption{\label{tab:cs} 
    Inclusive and exclusive cross sections to bound final states. 
    $(nlj)$ refers to the quantum numbers used in the calculation of the single-particle cross section, $\sigma_\text{sp}$, in the eikonal reaction theory.}
  \begin{ruledtabular}
    \begin{tabular}{rcrccr}
      $E$ (keV) 	        &  $J^\pi$ 	&  $\sigma_\text{exp}$ (mb) 	&  $(nlj)$ 	&  $\sigma_\text{sp}$ (mb) 	&  $C^2S_\text{exp}$\\
      \hline
      0	        & $3/2_1^-$	        & 12(4)	        & $2p_{3/2}$	& 21.7	& 0.55(17)\\  
      228	& ($1/2_1^-$)	        & 0.4(1)	& $2p_{1/2}$	& 20.8	& 0.019(4)\\
      320	& $7/2_1^-$	        & 43(3)	        & $1f_{7/2}$	& 14.3	& 3.00(21)\\
      977	& $7/2_2^-$	        & 1.2(6)	&               & 	& \\
      1155	& $3/2_2^-$(, $1/2^-$)	& 8.2(9)	& $2p_{3/2}$	& 18.7	& 0.44(5)\\
      1162	& $3/2_3^-$(, $1/2^-$)	& 5.3(3)	& $2p_{3/2}$	& 18.6	& 0.28(2)\\
      1209	& $3/2_4^-$(, $1/2^-$)	& 3.3(3)	& $2p_{3/2}$	& 18.5	& 0.18(1)\\
      1780	& 	                & 0.0(4)	&               & 	& \\
      1854	& 	                & 2.3(3)	&               & 	& \\
      2600	& $3/2_1^+$(, $5/2^+$)	& 8.8(4)	& $1d_{3/2}$	& 10.7	& 0.83(4)\\
      2628	& $7/2_3^-$(, $5/2^-$)	& 3.9(3)	& $1f_{7/2}$	& 12.2	& 0.32(3)\\
      \hline
      inclusive	& 	                & 91(4)	& 	& 	& \\
    \end{tabular}
  \end{ruledtabular}
\end{table}
The inclusive cross section to bound states in \nuc{43}{S} was determined from the number of particles identified in the S800 spectrograph and amounts to 91(4)~mb, slightly larger than but consistent with the previous measurement of the same reaction of 79(7)~mb~\cite{riley09}. The uncertainties include, in addition to statistical sources, the selection of the particle identification gate, the purity and intensity fluctuation of the incoming \nuc{44}{S} beam, uncertainties related to the transmission of the analysis line of the S800, and the thickness of the secondary target. The one-neutron reaction from \nuc{44}{S} was fully within the acceptance of the S800 spectrograph so that corrections were not necessary.

\begin{figure}[h]
\centering
\includegraphics[width=\columnwidth]{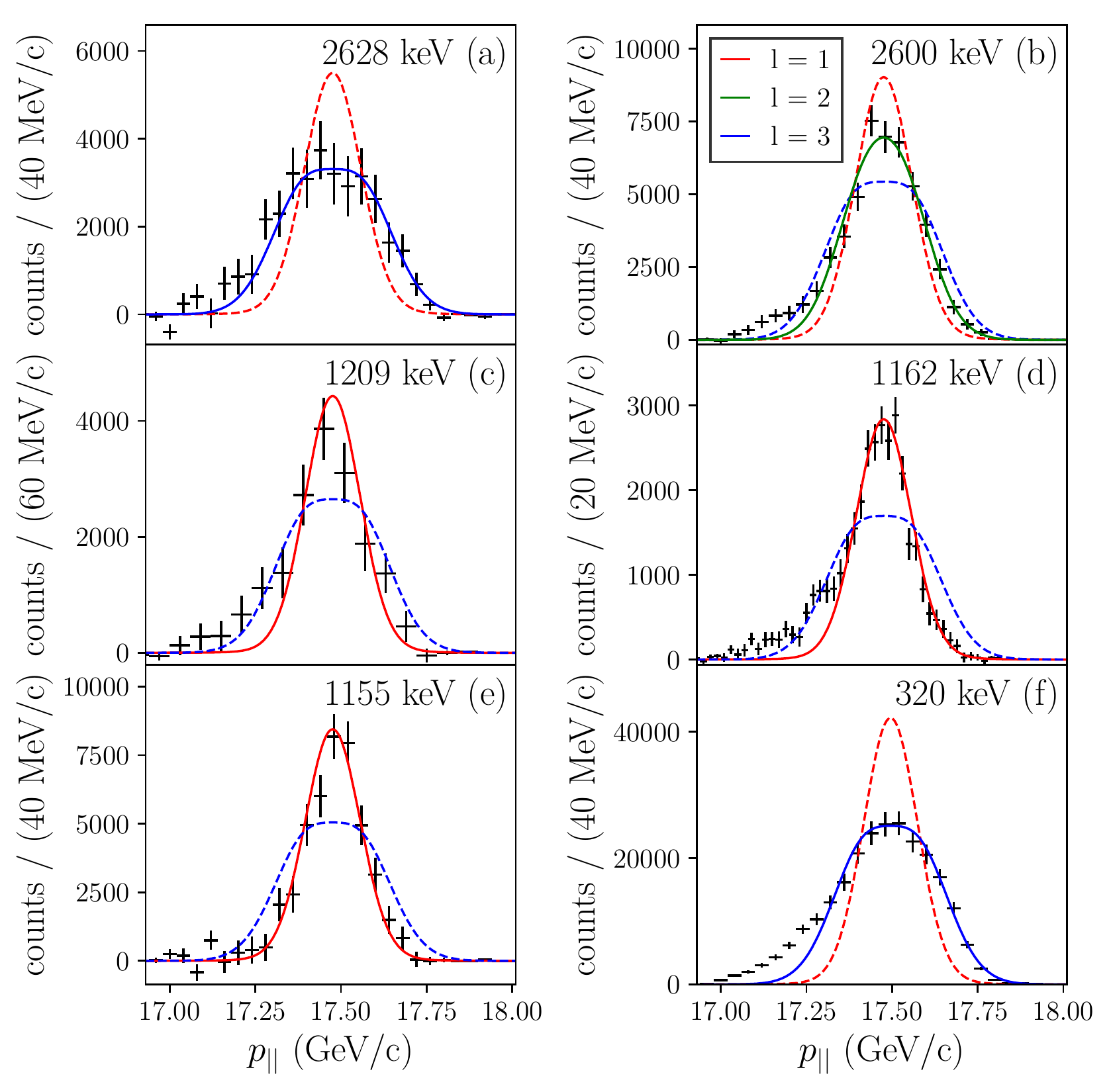}
\caption{Parallel momentum distributions of the one-neutron knockout reaction for several states in \nuc{43}{S}. Each panel shows the experimental parallel momentum distribution obtained by gating on \ghray\ transitions in black, compared to theoretical eikonal reaction model calculations for removal of a neutron from the $p$ (red), $d$ (green), and $f$ (blue) orbital. Panel (f) is the momentum distribution extracted in coincidence with the isomeric transition measured in IsoTagger and all others are obtained by gating prompt transitions measured in GRETINA.}
\label{fig:momdist}
\end{figure}

The parallel momentum distributions for several final states populated in \nuc{43}{S} are shown in Fig.~\ref{fig:momdist}.
In each case gates on the depopulating \ghray\ transitions were applied, and feeding from the higher-lying states was subtracted using the level scheme of \nuc{43}{S} and the efficiency of the \ghray\ detectors at the respective energies. The data are compared to theoretical calculations of neutron knockout from the $l=1$, 2, and 3 single-particle orbits using the eikonal reaction model~\cite{hansen03,tostevin14}. In this approach the projectile and target densities, taken from a Skyrme Hartree-Fock calculation for the projectile and assuming a Gaussian distribution for the light target, are used to construct the eikonal $S$ matrices for the ejectile- and nucleon-target interaction. The radial wave functions of the removed nucleon from each of the active orbitals are calculated in Woods-Saxon potentials with geometries constrained by the rms radius of the orbital from the Hartree-Fock calculation.
The calculated parallel momentum distributions were transformed into the laboratory system and folded with the experimental momentum resolution that was obtained from dedicated calibration runs. The theoretical calculations are normalized to the experimental counts in the 17.3 to 17.8~GeV/c momentum region. This momentum region was selected to eliminate the lower momentum tail region which is not reproduced by the eikonal reaction theory.
The states at 1155, 1162, and 1209~keV are well explained by neutron knockout from a $l=1$ $p$ orbital, probing the occupation of neutron orbits above the $N=28$ shell gap in the ground state of \nuc{44}{S}. On the other hand, the momentum distributions for the state at 2628 and the isomeric state at 320~keV are consistent with neutron knockout from the $l=3$ orbit. 
Thus, the spin-parity of the isomeric state of \nuc{43}{S}, already established as $7/2^-$~\cite{gaudefroy09}, is confirmed in the present work.
It is interesting to note that the \md\ of the state at 2600~keV, shown in Fig.~\ref{fig:momdist} (b), can only be reproduced by assuming removal of a neutron from the $1d_{3/2}$ orbital with $l=2$. This state is located close to the neutron separation energy~\cite{ringle09} and a candidate for a hole state in the $1d_{3/2}$ orbital below $N=20$. Such a state would not be populated in proton removal reactions in agreement with its non-observation~\cite{riley09,mijatovic18}. 
By subtracting the distributions of all excited states from the inclusive one, the \md\ and \cs\ directly populating the ground state of \nuc{43}{S} via one-neutron knockout reaction was extracted. Due to ambiguities in the level scheme and the unplaced prompt \grays, the distinction between the neutron knockout from the $f$ and $p$ orbits is less clear, but the momentum distribution is well described by knockout from the $2p_{3/2}$ orbital. In the following discussion, the spin-parity of the ground state of \nuc{43}{S} is assumed to be $3/2^-$, which was suggested from the transition rate from the $7/2^-$ isomeric state to the ground state~\cite{gaudefroy09}. 
The momentum distribution for the 977 and 1856~keV states are asymmetric and very broad, suggesting the population via a non-direct process. This would be expected from a collective rotational band member~\cite{ibbotson99, longfellow20}.

Using the eikonal reaction model calculations, the single-particle cross sections $\sigma_\text{sp}$ were calculated (see Table~\ref{tab:cs}). These depend on the effective separation energy for the final state and the quantum numbers of the orbital the nucleon was removed from. Using the spin and parity assignments shown in Fig.~\ref{fig:level} the spectroscopic factors, $C^2S_\text{exp} = \sigma_\text{exp}/\sigma_\text{sp}$, for each state were obtained. They are listed in Table~\ref{tab:cs}.

\section{Discussion}\label{sec:discussion}
The experimental level scheme is compared to the results of shell model calculations in Fig.~\ref{fig:level}. Two effective interactions in the full proton $sd$ and neutron $fp$ model space were used to calculate the excitation energies, transition probabilities, and spectroscopic factors. Effective charges $(e_\text{p}=1.35,e_\text{n}=0.35)$ and $g$ factors suggested in Ref.~\cite{chevrier14} have been used.
The SDPF-U~\cite{nowacki09} and SDPF-MU~\cite{utsuno12} interactions have been previously applied to \nuc{43,44}{S}~\cite{utsuno12, chevrier12, chevrier14, utsuno15, mijatovic18} and predict, at first glance, very similar level schemes shown in Fig.~\ref{fig:level}. The calculated energies, spectroscopic factors, and band assignments are listed in Table~\ref{tab:sm}.
\begin{table}
  \caption{\label{tab:sm} Results of the shell model calculations with the SDPF-U~\cite{nowacki09} and SDPF-MU~\cite{utsuno12} effective interactions. In addition to the bound states populated in the one-neutron knockout reaction, the members of rotational bands discussed in the text are listed.}
  \begin{ruledtabular}
    \begin{tabular}{rcrc|rcrc}
      \multicolumn{4}{c|}{SDPF-U} & \multicolumn{4}{c}{SDPF-MU} \\
      $E$ (keV) & $J^\pi$ & $C^2S$ & band & $E$ (keV) & $J^\pi$ & $C^2S$ & band \\
      \hline
         0 & $3/2_1^-$ & 0.64 & (a) &     0 & $3/2_1^-$ & 0.58 & (a) \\
       298 & $1/2_1^-$ & 0.15 & (a) &   134 & $1/2_1^-$ & 0.11 & (a) \\
       750 & $7/2_1^-$ & 2.66 & (b) &   601 & $7/2_1^-$ & 2.57 & (b) \\
      1010 & $7/2_2^-$ & 0.36 & (a) &   875 & $3/2_2^-$ & 0.40 & (c) \\
      1401 & $5/2_1^-$ & 0.01 & (a) &   935 & $7/2_2^-$ & 0.32 & (a) \\
      1405 & $3/2_2^-$ & 0.23 &     &  1035 & $5/2_1^-$ & 0.01 & (a) \\
      1990 & $5/2_2^-$ & 0.07 & (c) &  1444 & $5/2_2^-$ & 0.10 & (c) \\
      2053 & $1/2_2^-$ & 0.05 &     &  1665 & $1/2_2^-$ & 0.07 &     \\ 
      2132 & $5/2_3^-$ & 0.00 &     &  2143 & $9/2_1^-$ &      & (b) \\
      2366 & $9/2_1^-$ &      & (b) &  2196 & $3/2_3^-$ & 0.03 &     \\                
      2479 & $7/2_3^-$ & 1.87 & (c) &  2198 & $5/2_3^-$ & 0.04 &     \\ 
      2602 & $3/2_3^-$ & 0.01 &     &  2466 & $7/2_3^-$ & 1.46 & (c) \\
      3473 &$11/2_1^-$ &      & (b) &  2496 & $5/2_4^-$ & 0.08 &     \\ 
           &           &      &     &  2655 & $1/2_3^-$ & 0.01 &     \\ 
           &           &      &     &  2722 & $7/2_4^-$ & 0.51 &     \\ 
           &           &      &     &  3651 &$11/2_1^-$ &      & (b) \\
    \end{tabular}
  \end{ruledtabular}
\end{table}

In both cases, three rotational bands are predicted and these are labeled (a), (b), (c) in Table~\ref{tab:sm}. For the case of the SDPF-U interaction the band structure in \nuc{43,44}{S} is extensively discussed in Ref.~\cite{chevrier14}. The collective $7/2^-_2$ state at 977~keV is a member of the ground state band (a). Based on the comparison with the shell model calculation the 228~keV transition is a candidate for the decay from the $1/2^-_1$ state. In the shell model calculations, the $5/2^-_1$ state is predicted to decay to the $1/2^-_1$ state with a large $B(E2)$ value, but no such state was observed in the present work. The small cross sections for the $1/2^-_1$ and $7/2^-_2$ states suggest that they are not of single-particle character, in agreement with the calculations.

In the present work, we have for the first time identified a state built on top of the isomer. The state at 1854~keV decays to the $7/2^-$ isomer via the 1532~keV transition (see Fig.~\ref{fig:hodo}).
The 1856~keV transition has been tentatively assigned to a ground state decay. This would limit the spin and parity values to $J^\pi = (3/2,5/2,7/2)^-$. The momentum distribution for this state is rather broad, but no conclusion can be drawn. The state could be a candidate for the oblate $3/2^-$ band head predicted by the AMD calculations whose main decay branch is to the triaxial $7/2^-$ isomeric states~\cite{kimura13}. 
The shell model calculations do not predict a candidate for a corresponding state, but rather states with a $J^\pi = 7/2^-, 9/2^-, 11/2^-$ sequence (band (b)) are predicted, where the $9/2^-_1$ state is connected by strong $M1$ (0.24~$\mu_\text{N}$) and $E2$ (110~e$^2$fm$^4$ for the SDPF-MU interaction to the $7/2^-$ isomer. For SDPF-U the values are similar (see~\cite{chevrier14}). If the 1856~keV transition is placed elsewhere in the level scheme, the 1854~keV state could be a natural candidate for the $9/2^-_1$ state.
A firm spin and parity assignment for the 1854~keV state is required in order to draw further conclusions.
Finally, a third band-like structure is built on the $3/2^-_2$ state at 1155~keV. The $7/2^-_3$ state at 2628~keV decays to the state at 1780~keV via the 850~keV transition, as well as to the 1155 or 1162~keV state by emission of a 1469~keV \gray. The 1780~keV state is not populated directly, it decays via the 625~keV transition, a likely spin assignment is thus $5/2^-$. The $3/2^-$ and $7/2^-$ states can be associated with the shell model states at 1405 (875) and 2479 (2466) or 3093 (2722)~keV in the SDPF-U (SDPF-MU) results, based on the comparison of the spectroscopic factors. However, none of the shell model states shows a decay pattern similar to the experimentally observed one. The decay of these is fragmented to several states below with individual $B(E2)$ values around 1-100~e$^2$fm$^4$. The $7/2^-_3$ member of the second prolate band (c) at 2479~keV predicted by the SDPF-U calculations~\cite{chevrier14}, for example, has a strong $B(E2)$ value for the decay to the $5/2^-_2$ band head (1990~keV), but the predicted branching ratio is only 15.2~\% owing to the higher energy difference for the other possible decays to lower lying states.
Furthermore, the SDPF-MU calculations, in contrast with those using SDPF-U, predict a strong transition from the $5/2^-_2$ state to the $3/2^-_2$ state suggesting a $3/2^-$ band head instead, more in line with the results from the AMD calculations~\cite{kimura13}. Clearly, more experimental investigation is required to establish the band structure and determine its deformation characteristics.

The neutron knockout cross sections to the bound, shell-model final states in \nuc{43}{S} have been calculated using the theoretical spectroscopic factors $C^2S$ and the single-particle cross sections $\sigma_\text{sp}$
\begin{equation}
\sigma(J^\pi) = \left(\frac{A}{A-1}\right)^N C^2S(J^\pi) \sigma_\text{sp}(nlj,S_\text{n}+E(J^\pi)). \nonumber
\label{eq:cs}
\end{equation}
They are compared to the experimental results in Fig.~\ref{fig:level}.
The inclusive theoretical cross section was calculated by summing the contributions of all states up to the experimental neutron separation energy $S_\text{n}(^{43}\text{S}) = 2629$~keV~\cite{ringle09}. The inclusive cross section amounts to 94.3 (91.7)~mb for the SDPF-U (SDPF-MU) interactions. Experimentally the cross section populating positive parity states by $sd$-shell neutron removal, which are outside of the model space of the calculations, amounts to at least 8.8(4)~mb. An estimate for the reduction factor $R_\text{S}$~\cite{gade08,tostevin14} is thus given by the ratio of the cross section to $fp$ states to the theoretical value and amounts to 0.87(0.90) for the two effective interactions, in line with the systematics~\cite{gade08,tostevin14}.

The isomeric $7/2_1$ state carries the major fraction of the single-particle strength, but still significantly less than expected from a pure $\nu(f_{7/2})^{-1}$ configuration. This is in agreement with the interpretation of the electric quadrupole moment of this state~\cite{chevrier12}, which is significantly larger than expected for a single hole in the $1f_{7/2}$ orbital. The shell model calculations predict that a large fraction of the $1f_{7/2}$ strength is located close to the neutron separation energy. Experimentally, the strength to unbound states is inaccessible in the present setup, therefore, part of the $1f_{7/2}$ strength could be missed in the experiment. Three states with significant $l=1$ strength are observed around 1200 keV. This is not reproduced by the shell model calculations which predict only one excited $3/2^-$ state in this energy region. The 1162~keV state decays to the 977~keV $J^\pi = 7/2^-$ state. The lifetime of this state, if the present level ordering is adopted, amounts to 15(2)~ps~\cite{mijatovic18}. Such a state is not found in the shell model calculations. 
The spectroscopic factors for $1/2^-$ and $5/2^-$ states are small as it is expected that the occupation of the $2p_{1/2}$ and $1f_{5/2}$ orbitals in the ground state of \nuc{44}{S} is small. 

The spectroscopic factors can also be compared to the $N=28$ isotones. \nuc{47}{Ca} has been studied in detail by pickup transfer reactions using \nuc{48}{Ca} targets. The spectroscopic factor for the ground state amounts to $C^2S=6.22$~\cite{williamsnorton77}. Using the typical reduction $R\approx 0.7$ of the spectroscopic strength when comparing to the shell model and assuming this reduction is applicable to each transition, this compares well with the expectation of the independent particle model of $C^2S=8$ for the $1f_{7/2}$ orbital. For the $3/2^-$ state at 2014~keV only a small spectroscopic factor of $C^2S=0.1$ was found. 
For the radioactive \nuc{45}{Ar} nucleus a measurement of the neutron knockout reaction from \nuc{46}{Ar} also found a small spectroscopic factor for the first excited $3/2^-_1$ state of $0.2(2)$~\cite{gade05}. In the same experiment, the ground state was populated with a spectroscopic factor of $C^2S=4.9(7)$. These values are in qualitative agreement with the shell model calculations which predict spectroscopic factors of 0.59 and 5.34 for the SDPF-U and 0.79 and 5.00 for the SDPF-MU interactions. The results indicate that $N=28$ is a good shell closure in Ca and Ar nuclei.

In the present work, the spectroscopic strength for the population of the first $7/2^-$ state amounts to 3.00(21), significantly lower than for the heavier isotones. The spectroscopic factor for the $3/2^-$ ground state is 0.55(17). However, several other states are populated by the removal of a neutron from the $p$ orbitals, as evidenced from the momentum distributions shown in Fig.~\ref{fig:momdist}. While the present experiment cannot distinguish between removal of a $2p_{3/2}$ and a $2p_{1/2}$ neutron, the latter is unlikely as the $2p_{1/2}$ is expected to lie higher in energy. The shell model calculations also do not predict large \sfs\ for the $J^\pi = 1/2^-$ states (see Table~\ref{tab:sm}). The observed fragmentation of the $2p_{3/2}$ strength is not predicted by the shell model calculations.  If the experimental spectroscopic factors for the $2p$ states are added, and normalized using the reduction factor~\cite{gade08,tostevin14}, $R_\text{S}$, as determined from other nuclei as a function of the separation energies, the summed normalized spectroscopic strength can be used as a indicator for the occupation number. In the present case the sum amounts to 1.8(4) where the uncertainty is dominated by a systematic uncertainty of $R_\text{S}$ which has been assumed to be 20\%. This suggests that the ground-state configuration of \nuc{44}{S} is composed of up to two neutrons in the $2p_{3/2}$ orbital. The shell model calculation for the summed spectroscopic strengths amount to $\sum C^2S(1f_{7/2}) = 4.89$ (4.35) and $\sum C^2S(2p_{3/2}) = 0.89$ (1.01) for the SDPF-U (SDPF-MU) interactions. The occupation numbers for the $2p_{3/2}$ orbital in the ground state of \nuc{44}{S} are 1.18 and 1.38, respectively.
If the cross-shell $\pi sd - \nu fp$ tensor component of the SDPF-MU matrix elements is removed, the summed spectroscopic strength, up to $S_\text{n}$, increases to 5.26 for the $1f_{7/2}$ orbital. This is in line with the interpretation that the proton-neutron tensor interaction is driving the shell evolution in this exotic region of the nuclear chart~\cite{utsuno12}.

The location of the $7/2^-$ and $3/2^-$ states in \nuc{43}{S} already suggested the inversion of the normal $(1f_{7/2})^{-1}$ and intruder $(2p_{3/2})$ neutron configurations. The present experiment proves for the first time an intruder dominance of the ground state of \nuc{44}{S}. This is significantly different from the less exotic isotones, and the increase in $2p_{3/2}$ configurations in the ground state of \nuc{44}{S} compared to \nuc{46}{Ar} is abrupt. In the even more exotic isotone \nuc{41}{Si} only one transition was observed~\cite{sohler11}, however, many more low-lying states are expected based on the shell model calculations. \nuc{41}{Si} would be an ideal testing ground for the shell model calculations, since there the SDPF-U(-SI) and SDPF-MU interactions predict very different spectroscopic factors for the one-neutron removal reaction from \nuc{42}{Si}.

Finally, in the present experiment the population of a positive parity state at 2600~keV was observed. The spectroscopic factor amounts to 0.83(4) assuming removal of a neutron from the $1d_{3/2}$ orbital. This value can be compared to the isotone \nuc{47}{Ca}, where the $3/2^+_1$ state is located at 2580 keV and has a deduced \sf\ of 1.18~\cite{williamsnorton77}, determined from the $(d,t)$ transfer reaction measurement. This state lies outside of the model space and is not described with the present shell model calculations.




\section{Summary and outlook}\label{sec:summary}
In summary, we have performed spectroscopy of \nuc{43}{S} using the one-neutron knockout reaction from \nuc{44}{S}. Using prompt and delayed \ghray\ spectroscopy in coincidence, the level scheme of \nuc{43}{S} was constructed. Previously, this was beset with ambiguities due to the presence of a long-lived isomeric state in \nuc{43}{S}. Final-state exclusive momentum distributions of the residue allowed for firm spin and parity assignments. The level ordering and assignments of a recent lifetime measurement~\cite{mijatovic18} were revised.
A state above the isomer was identified for the first time, but its properties could not be reproduced using shell model calculations. Coulomb excitation measurements using an isomeric \nuc{43}{S} beam could help in resolving this issue.
A band-like structure built on a $3/2^-$ state was observed, but further experimental investigation is required to confirm a band and determine its properties.
The cross sections for the population of states originating from the removal of a $2p_{3/2}$ neutron from the \nuc{44}{S} ground state were found to be large. This is a direct measure of the amount of intruder configuration in the ground state of \nuc{44}{S} and quantifies the $N=28$ shell quenching in this exotic nucleus.

\acknowledgments
We would like to thank the NSCL staff for the preparation of the radioactive beam at the Coupled Cyclotron Facility. This work was supported by the U.S National Science Foundation under Grant No. PHY-1306297, PHY-1102511, and PHY-1565546, by the U.S. Department of Energy (DOE) National Nuclear Security Administration through the Nuclear Science and Security Consortium under Award No. DE-NA0003180, and by the DOE-SC Office of Nuclear Physics under Grants No. DE-SC002045. GRETINA was funded by the DOE, Office of Science. Operation of the array at NSCL was supported by the DOE under Grants No. DE-SC0014537 (NSCL) and No. DE-AC02-05CH11231 (LBNL).
SM acknowledges support from JSPS Grant-in-Aid for JSPS Research Fellow Grant Number JP15J08882. 
KW acknowledges support from the Spanish Ministerio de Econom\'ia y Competitividad RYC-2017-22007.
JAT acknowledges support from the Science and Technology Facilities Council (U.K.) Grant No. ST/L005743/1.
\bibliography{arxiv}

\end{document}